\documentclass[reprint,superscriptaddress,aps,prb]{revtex4-2}

\usepackage{amsmath}
\usepackage{bm}
\usepackage{tabularx}
\usepackage{comment}
\usepackage{here}
\usepackage{dcolumn}
\usepackage{float}
\usepackage{stfloats}
\usepackage{graphicx}
\usepackage{amssymb}
\usepackage{booktabs} 
\usepackage{lineno}

\begin{document}
\newcommand{\tn}{$T_{\rm N}$}
\newcommand{\tord}{$T_{\rm o}$}
\newcommand{\mub}{${\mu}_{B}$}

\title{Successive Phase Transitions in the Quasi-Kagome Lattice System URhSn Studied by Resonant X-ray Scattering}

\author{Chihiro Tabata}
\email{tabata.chihiro@jaea.go.jp}
\affiliation{Materials Sciences Research Center, Japan Atomic Energy Agency, Tokai 319-1106, Japan}
\affiliation{Advanced Science Research Center, Japan Atomic Energy Agency, Tokai 319-1106, Japan}

\author{Fusako Kon} 
\affiliation{Graduate School of Science, Hokkaido University, Sapporo 080-0810, Japan}

\author{Ruo Hibino}
\altaffiliation[Present address: ]{Department of Physics, Kobe University, Kobe, Hyogo 657-8501, Japan}
\affiliation{Graduate School of Science, Hokkaido University, Sapporo 080-0810, Japan}

\author{Yusei Shimizu}
\altaffiliation[Present address: ]{Institute for Solid State Physics (ISSP), The University of Tokyo, Kashiwa, Chiba 277-8581, Japan}
\affiliation{Institute for Materials Research, Tohoku University, Oarai  311-1313, Japan}

\author{Hiroshi Amitsuka}
\affiliation{Graduate School of Science, Hokkaido University, Sapporo 080-0810, Japan}

\author{Koji Kaneko}
\affiliation{Materials Sciences Research Center, Japan Atomic Energy Agency, Tokai 319-1106, Japan}
\affiliation{Advanced Science Research Center, Japan Atomic Energy Agency, Tokai 319-1106, Japan}

\author{Yoshiya Homma}
\affiliation{Institute for Materials Research, Tohoku University, Oarai  311-1313, Japan}

\author{Dai Aoki}
\affiliation{Institute for Materials Research, Tohoku University, Oarai  311-1313, Japan}

\author{Hironori Nakao}
\affiliation{Photon Factory, Institute of Materials Structure Science, High Energy Accelerator Organization, Tsukuba 305-0801, Japan}

\date{\today}

\begin{abstract}
Successive phase transitions in the quasi-kagome compound URhSn were investigated by resonant X-ray scattering (RXS) at the uranium $M_4$ edge. 
In the high-temperature phase between 16 K and 54 K, an additional RXS signal was detected superposed onto fundamental reflections in both $\pi$-$\sigma'$ and $\pi$-$\pi'$ polarization channels. 
Upon cooling below 16 K, reported as a ferromagnetic phase along $c$, substantial enhancements were observed again in the both polarization channels at the 300 reflection, demonstrating a simultaneous emergence of in-plane spin alongside the $c$-axis ferromagnetic components. 
The observed behavior can be interpreted by an antiferro-quadrupole (AFQ) order of $O_{yz}$ or $O_{zx}$ characterized by a propagation vector $\bm{q} = 0$ in the intermediate phase, which then coexists with a ferromagnetic component below 16 K. 
The resulting ground state structure breaks the mirror symmetry perpendicular to the kagome plane, identifying the formation of a unique AFQ order with either chirality or polarity in URhSn.

\end{abstract}

\maketitle

Magnetic frustration has been recognized as a key determinant in the emergence of unconventional state of matter.
The kagome lattice, characterized by a corner-sharing triangular network, involves highly frustrated correlations as well as characteristic band structures \cite{Park2025}.
These features give rise to exotic electronic states, such as spin liquids \cite{Savary2017,Zhou2017,Isono2016}, and recently discovered coexistence of chiral charge order with superconductivity \cite{Jiang2021,Shumiya2021,Wang2021}. 
Whereas these findings have been established for spin-based $d$-electron compounds, $f$-electron systems with relevant spin-orbit interactions have also come into scope of intensive research.

The ternary intermetallic $MTX$ series ($M$: lanthanides and actinides, $T$: transition metals, $X$: $p$-block elements such as Al, Sn, Ge, etc.), which adopts the ZrNiAl-type structure ($P\bar{6}2m$; no. 189, $D_{3h}^{3}$), is known as a distorted two-dimensional kagome system, providing a platform for studying frustrated f-electron systems \cite{Gupta2015}.
Indeed, $R$AgGe compounds ($R = $ Tb-Yb) exhibit complex magnetic phase diagrams and multistep metamagnetic transitions owing to geometrical frustration \cite{Morosan2004,Zhao2016,Mochidzuki2017,Schmiedeshoff2011}.
One of the members, HoAgGe, generated considerable interest due to its multiple magnetic phases regarded as a naturally occurring kagome spin-ice state \cite{Li2022}.

This study focuses on the 5\textit{f}-electron-based quasi-kagome system URhSn.
Early studies on polycrystalline URhSn reported two successive phase transitions at $T_{\rm C}$ = 16 K and $T_{\rm o}$ = 54 K \cite{Palstra1987,Tran1991,Tran1995}.
The low-temperature phase (low-$T$ phase) below $T_{\rm C}$ has been considered a ferromagnetic (FM) state.
On the other hand, the order parameter (OP) of the high-temperature phase (high-$T$ phase) between $T_{\rm C}$ and $T_{\rm o}$ remains unidentified.
$^{119}$Sn M\"{o}ssbauer spectroscopy and Sn-NMR have suggested the absence of an additional internal field \cite{Kruk1997,Tokunaga2023}, and neutron powder diffraction detected no superlattice reflections in this high-$T$ phase \cite{Mirambet1995}, suggesting a nonmagnetic order.
In contrast, a recent study on the precise thermodynamic and transport properties of single-crystalline URhSn revealed a second-order phase transition accompanied by a substantial release of the 5\textit{f} electronic entropy of $R$ln3 \cite{Shimizu2020}. 
Moreover, $T_{\rm o}$ increases under a magnetic field applied along the $c$-axis, which is atypical for a simple antiferromagnetic (AFM) order but is often observed in systems with an electric quadrupole order \cite{Effantin1985,Shiina1998,Tayama2001,Kaneko2003,Indoh2004}.

In this study, the successive phase transitions of URhSn were investigated via resonant X-ray scattering (RXS), a technique highly sensitive to quadrupole order.
A resonance at the uranium $M_4$ absorption edge is dominated by an electric dipole-dipole ($E1$-$E1$) transition between the inner-core 3$d$ levels and the unoccupied 5\textit{f} levels, providing high sensitivity to 5\textit{f}-electron order. 

Single-crystalline URhSn was grown using the Czochralski method as described in Ref. ~[\citenum{Shimizu2020}], and shaped into a 0.5 $\times$ 1 $\times$ 1 mm$^3$ rectangular parallelepiped with a polished ($10\bar{1}0$) surface. 
The RXS experiments were performed at BL-11B of Photon Factory at KEK using a two-circle diffractometer for soft X-rays installed in a high-vacuum chamber \cite{Nakao2014}. 
The sample was mounted on a Cu block with silver paste, and was cooled to 10~K using a liquid $^4$He flow cryostat.
Two experimental configurations were employed for the diffraction measurements: the $(hk0)$ and $(h0l)$ horizontal scattering planes for configurations 1 and 2, respectively.
The incident X-ray beam was $\pi$-polarized, and the polarization of the scattered X-ray was analyzed using the (111) reflection of an Al crystal.
This Al(111) analyzer yields a diffraction angle of 90.8$^{\circ}$ for an X-ray of 3.72 keV, with a calculated polarization separation efficiency of 99.8\%.
A silicon drift detector was placed downstream of the analyzer crystal.

\begin{figure}[h]
\centering
        \includegraphics[width=6.0cm]{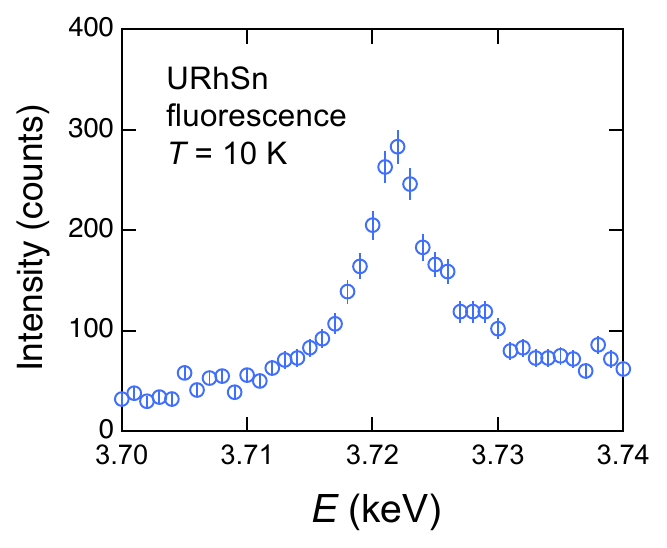}
    \caption{(Color online) Fluorescence spectrum of the $M_{4}$ edge of URhSn.}
    \label{fig:Escan}
\end{figure}

First, the $M_4$ absorption edge of U in URhSn was determined from the fluorescence spectrum as shown in Fig. \ref{fig:Escan}.
A clear peak was observed at 3.72 keV, and this was used as a resonance energy for subsequent measurements.
Using X-rays of this energy, the reciprocal space was surveyed by line scans along the principal directions.
With the $(h0l)$ horizontal scattering plane in configuration 2, scans along [100] and [001] were performed around (3,0,0).
Configuration 1, employing the $(hk0)$ scattering plane, was used to scan along [110], from (1,1,0) to (2,2,0).
In addition, a mesh scan was carried out in configuration 2 to survey the $(h0l)$ plane. 
Within these explored $Q$-spaces, no superlattice reflection was observed at 30 K, below $T_{\rm o}$.

\begin{figure}[h]
\centering
        \includegraphics[width=6.0cm]{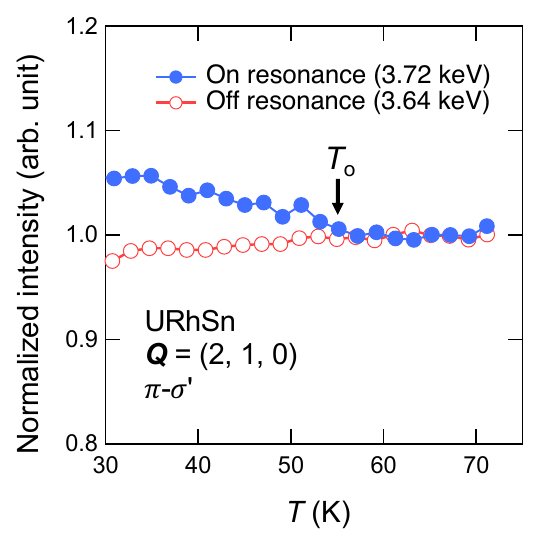}
    \caption{(Color online) Temperature dependence of the RXS intensity of 210 reflection at the resonance condition ($E = 3.72$ keV) and the nonresonant condition ($E = 3.64$ keV), measured in configuration 1. The intensity is normalized by the background signal above 60 K.}
    \label{fig:reso}
\end{figure}

Instead of superlattice reflections, an intensity increase was identified at fundamental Bragg reflections. 
Figure \ref{fig:reso} shows the temperature dependence of the relative intensity of the 210 Bragg reflection in the $\pi$-$\sigma'$ channel measured under the on- and off-resonance conditions at 3.72 and 3.64 keV, respectively. 
As the temperature decreases, the intensity starts to develop below $T_{\rm o}$ only under the on-resonance condition.
Although this behavior is indicative of resonant enhancement, the effect was too weak to be resolved as a distinct resonant peak in the energy dependence (Fig. S1\cite{Suppliment}).
Nevertheless, this resonant enhancement, revealed through careful temperature-dependent measurements, suggests the onset of a uranium 5\textit{f}-electron order characterized by a propagation vector $\bm{q} = 0$.

\begin{figure}[ht]
    \begin{center}
        \includegraphics[width=8.0cm]{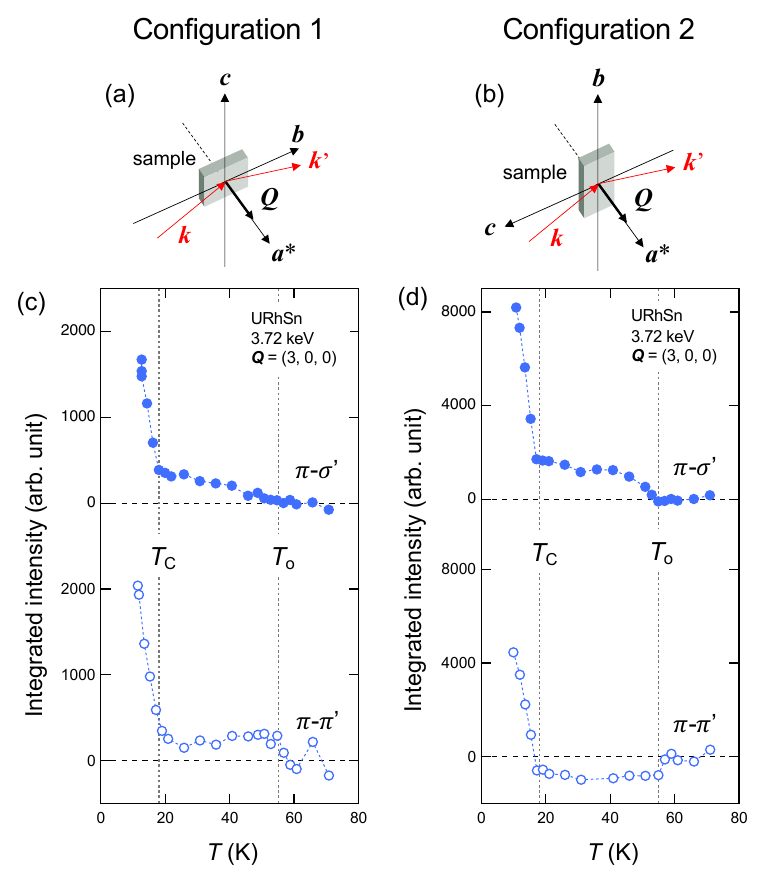}
    \caption{(Color online) Temperature dependence of the RXS intensity of 300 Bragg reflection at the resonance condition ($E = 3.72$ keV) for two different configurations. The intensities are shown after subtracting the signals measured in the non-ordered phase above $T_{\rm o}$.}
    \label{fig:pol}
    \end{center}
\end{figure}

To clarify the origin of this $\bm{q} = 0$ intensity enhancement, the polarization dependence of the resonant signal at the 300 reflection was investigated using two experimental configurations as shown in Fig.~\ref{fig:pol} (a) and (b).
In configuration 1, the $c$-axis of the sample was oriented perpendicular to the horizontal scattering plane, whereas in configuration 2, it was aligned within the scattering plane.

As shown in Fig.~\ref{fig:pol}(c), the integrated intensity of the 300 Bragg reflection at 3.72~keV shows a gradual increase in the $\pi$-$\sigma'$ channel below $T_{\rm o}$ with decreasing temperature. 
The $\pi$-$\pi'$ scattering also slightly increases below $T_{\rm o}$, although it is less distinct than that in the $\pi$-$\sigma'$ channel due to a large background arising from an overlapping fundamental reflection.
The onset of the order at $T_{\rm o}$ was more clearly observed in configuration 2; the $\pi$-$\sigma'$ component exhibited a clear upturn, 
whereas the intensity in the $\pi$-$\pi'$ channel decreases below $T_{\rm o}$.
In the low-$T$ phase, all four data sets exhibit a significant increase in intensity below $T_{\rm C}$.
Consequently, the intensities in both $\pi$-$\sigma'$ and $\pi$-$\pi'$ channels for both configurations are sensitive to the both transitions at $T_{\rm C}$ and $T_{\rm o}$.

Hereafter, the possible OPs for each ordered phase are discussed.
Based on the atomic model by Hannon et al.\cite{Hannon1988}, the polarization-dependent RXS amplitude for the $E1$-$E1$ transitions is expressed as:
\begin{equation}
f_{\rm RXS} = F^{(0)} \bm{\varepsilon}' \cdot \bm{\varepsilon}
            - i F^{(1)} (\bm{\varepsilon}' \times \bm{\varepsilon}) \cdot \bm{u}
            + F^{(2)} (\bm{\varepsilon}' \cdot \bm{u})(\bm{\varepsilon} \cdot \bm{u}).
\label{eq:F}
\end{equation}
The first, second, and third terms correspond to the contributions from the charge, magnetic dipole, and electric quadrupole, respectively.
The vectors $\bm{\varepsilon}$ and $\bm{\varepsilon'}$ are the unit vectors of the incident and scattered photon polarizations, and $\bm{u}$ is the unit vector associated with the local uniaxial field induced by the magnetic dipole and the electric quadrupole.
$F^{(i)}$ are complex coefficients related to the inner process of the resonance.

\begin{figure}[ht]
\centering
        \includegraphics[width=7cm]{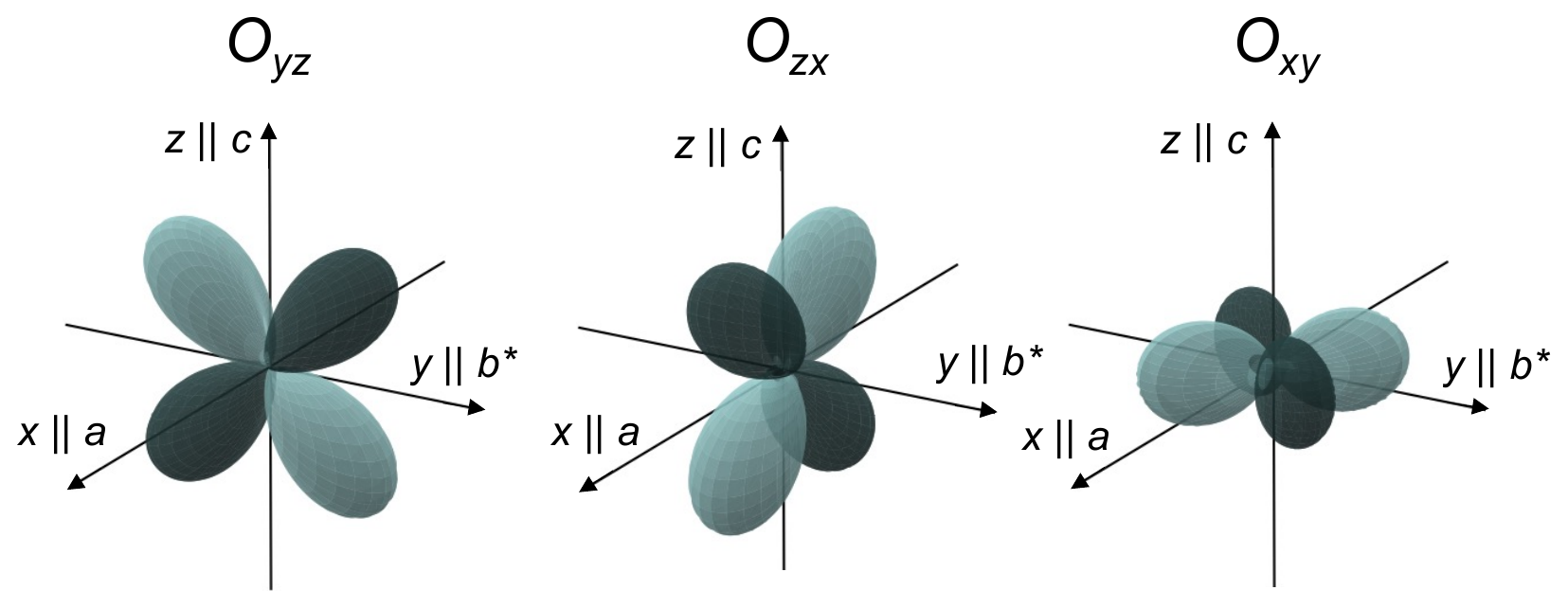}
    \caption{(Color online) Schematic illustration of the quadrupoles considered as candidate order parameters in the intermediate phase of URhSn. The $O_{20}$ component is not shown. The quadrupoles at the crystallographic site ($x_{\rm U}$, 0, 0) are shown as a representative example. The other two sites in the unit cell are generated by applying the threefold rotational symmetry operation. The $a$, $b^*$, and $c$ axes correspond to [$2\bar{1}\bar{1}0$], [$01\bar{1}0$], and [0001] directions, respectively.}
    \label{fig:xyz}
\end{figure}

Based on Eq. \ref{eq:F} and the observation, possible OPs for the two transitions at $T_{\rm o}$ and $T_{\rm C}$ of URhSn were considered.
First, charge order is straightforwardly excluded because charge scattering does not contribute to $\pi$-$\sigma'$ scattering, as indicated by the first term of Eq. \ref{eq:F}. 
For U sites in URhSn with $C_{2v}$ symmetry, the propagation vector of $\bm{q} = 0$ allows three magnetic dipole OPs, $J_x$, $J_y$, $J_z$, and four quadrupole OPs, $O_{20}$, $O_{yz}$, $O_{zx}$, and $O_{xy}$.
These OPs are classified according to their non-zero contribution to RXS amplitude in terms of polarization channel and configurations, as summarized in Table \ref{tab:OP}.
Here, the crystallographic axes $a$, $b^*$, and $c$ are defined to correspond to the $x$-, $y$-, and $z$-axes, respectively.
The quadrupole OPs shown in Fig. \ref{fig:xyz} are represented according to this definition.

\renewcommand{\arraystretch}{1.4}
\tabcolsep=5mm
\begin{table}[htb]
\centering
\caption{The detectable order parameters of magnetic dipoles (MD) and electric quadrupoles (EQ) for each experimental configuration.}
\begin{tabular}{ccc}
\hline \hline
\multicolumn{3}{c}{Configuration 1}
              \\ \hline
$\varepsilon$-$\varepsilon'$ & MD  & EQ   \\   
\midrule
$\pi$-$\sigma'$   & $J_x$, $J_y$        & $O_{yz}$, $O_{zx}$         \\
$\pi$-$\pi'$    &  $J_z$        & $O_{yz}$, $O_{zx}$, $O_{xy}$ \\ 
\hline \hline
\multicolumn{3}{c}{Configuration 2}
 \\ \hline
 $\varepsilon$-$\varepsilon'$ & MD  & EQ   \\     
\midrule        
$\pi$-$\sigma'$    & $J_x$, $J_y$, $J_z$ & $O_{yz}$, $O_{zx}$, $O_{xy}$ \\
$\pi$-$\pi'$     &  $J_x$, $J_y$ & $O_{20}$, $O_{yz}$, $O_{zx}$, $O_{xy}$ \\ 
\hline \hline
\end{tabular}
\label{tab:OP}
\end{table}

Concerning magnetic dipoles, the in-plane ($J_x$, $J_y$) and the out-of-plane ($J_z$) exhibit different polarization dependences;
in configuration 1, $\pi$-$\sigma'$ and $\pi$-$\pi'$ reflect the in-plane and out-of-plane components, respectively.
On the other hand in configuration 2, $\pi$-$\pi'$ is not sensitive to $J_z$, whereas $\pi$-$\sigma'$ detects all three components.
Similar characterization is applicable to electric quadrupoles.
$O_{yz}$, $O_{zx}$ contribute to all four configurations.
In contrast, $O_{xy}$, does not give signal in $\pi$-$\sigma'$ of the configuration 1, 
whereas $O_{20}$ can be detected only in $\pi$-$\pi'$ of the configuration 2.

Using the classification in Table \ref{tab:OP}, the candidates for the OPs in each phase were narrowed down.
At the transition temperature $T_{\rm o}$, intensity variations were observed in all setups (i.e., both the $\pi$-$\sigma'$ and $\pi$-$\pi'$ channels in each configuration), regardless of whether the signal intensity increased or decreased \footnote{It is not trivial whether the signal intensity increases or decreases, as it depends on detailed interference with non-resonant scattering --including contributions from Rh and Sn-- which occurs in the $\bm{q} = 0$ case.}.
The OPs consistent with these observations are:
\begin{itemize}
    \item Magnetic dipoles with both out-of-plane ($J_z$) and in-plane ($J_x$ and $J_y$) components.
    \item Electric quadrupoles with $z$-components ($O_{yz}$ and $O_{zx}$).
\end{itemize}
Since the magnetic dipole order has been excluded by previous studies using NMR, M\"{o}ssbauer spectroscopy, and neutron diffraction \cite{Kruk1997,Tokunaga2023, Mirambet1995}, the remaining candidates are non-magnetic OPs, specifically $O_{yz}$ or $O_{zx}$. 

In the low-$T$ phase, signals measured in all setups, including $\pi$-$\sigma'$ and $\pi$-$\pi'$ scattering in both configurations, exhibited a pronounced upturn below $T_{\rm C}$.
Given the substantial ferromagnetic component observed in previous magnetization and neutron diffraction measurements \cite{Tran1991,Shimizu2020,Mirambet1995}, a magnetic ordering occurs.
The reported magnetic  order with $J_z$ alone, however, does not satisfy the observed enhancement.
Since the $O_{yz}$ or $O_{zx}$ order is already established in the high-$T$ phase, the magnetic structure below $T_{\rm C}$ must be symmetry-compatible with the underlying quadrupole order, thereby allowing the induction of in-plane components, $J_x$ and $J_y$.
These RXS observations can be naturally explained by a canted magnetic order with finite components of $J_x$, $J_y$, and $J_z$ as follows:
\begin{itemize}
    \item The large enhancement in the $\pi$-$\pi'$ channel in configuration 1 is attributed to the out-of-plane ferromagnetic component, $J_z$.
    \item The $\pi$-$\sigma'$ scattering in configuration 1 and the $\pi$-$\pi'$ scattering in configuration 2 are ascribed to the in-plane components, $J_x$ or $J_y$.
\end{itemize}

\begin{figure}[h]
        \includegraphics[width=8.0cm]{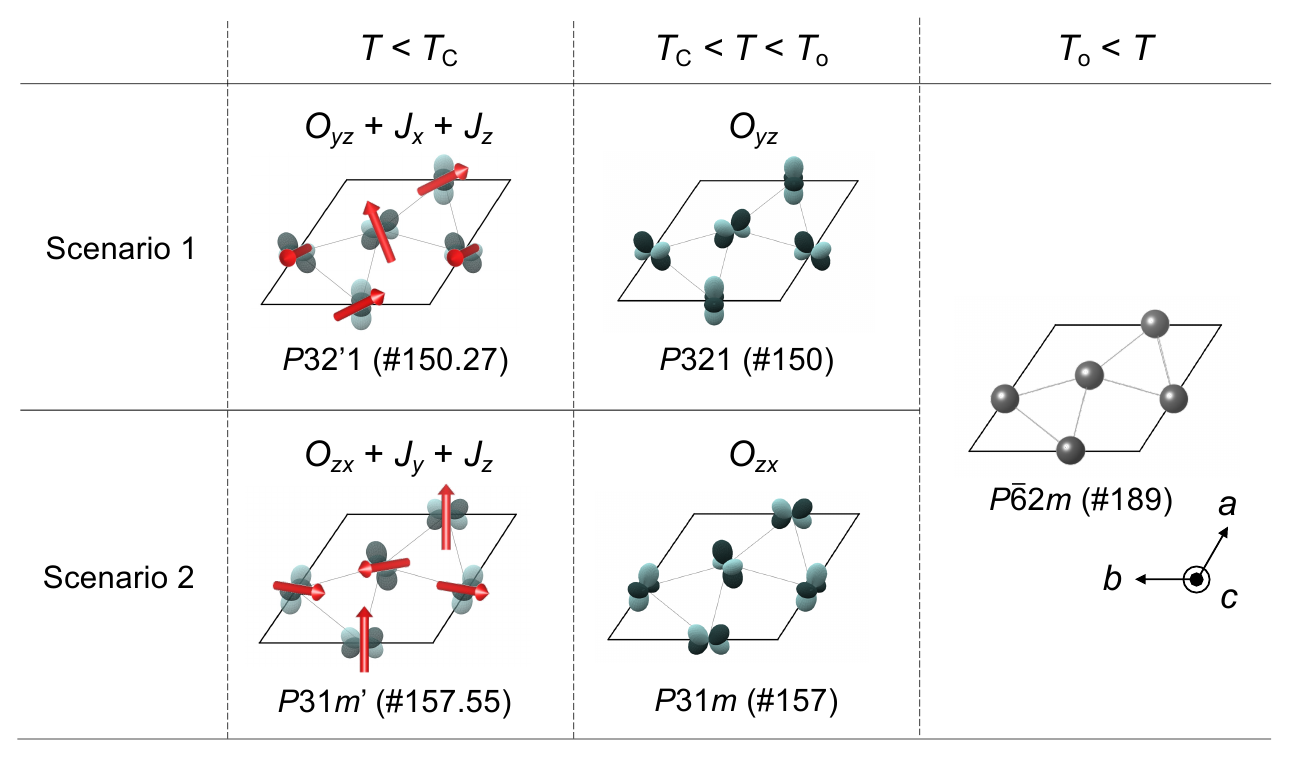}
    \caption{(Color online) Two possible scenarios for the symmetry lowering at the successive phase transitions of URhSn. Rh and Sn sites are not shown.}
    \label{fig:scenario}
\end{figure}

Consequently, two possible scenarios for the successive phase transitions in URhSn are proposed, which are schematically depicted in Fig.\ref{fig:scenario}.
The main difference between these two scenarios lies in responsible quadrupoles, either $O_{yz}$ or $O_{zx}$.
In both scenarios, the OPs of the high-$T$ phase break the sixfold rotoinversion axis of the parent space group, leaving the threefold rotational axis parallel to the $c$-axis.
Additionally, the mirror symmetries $\sigma_{zx}$ and $\sigma_{xy}$ at the U site are lost in the case of $O_{yz}$ (scenario 1).
This reduces the system's symmetry to $P321$ (no. 150), which belongs to the Sohncke groups, allowing a chiral structure \cite{Harima2023}. 
On the other hand, in the $O_{zx}$ case (scenario 2), the two-fold rotational axis along the $a$-axis is lost, resulting in the space group $P31m$ (no. 157).
The space group $P31m$ belongs to the polar space groups.
Polarity along the $c$-axis can be imparted to the system by freeing $z$ coordinates of the U, Rh, and Sn sites, which can be examined by crystal structure analysis \cite{Harima2023}.
The chiral and polar pairs are schematically shown in Fig. S2 \cite{Suppliment}.

The present qualitative analysis of the RXS signal does not allow us to draw a definitive conclusion for the OP of URhSn.
While the azimuthal angle dependence in the $\pi$-$\sigma'$ channel could, in principle, make it possible to distinguish between $O_{yz}$ and $O_{zx}$, contamination from the fundamental reflection intensity due to the imperfect polarization makes it difficult to obtain reliable intensities in the present case with $\bm{q}=0$.
One approach to distinguishing between two scenarios is to investigate the low-$T$ phase using other magnetic probes such as neutron scattering.
As mentioned above, the low-$T$ phase involves an in-plane dipole component. 
Reflecting the underlying quadrupole order of the high-$T$ phase, the subsequent magnetic structure in the low-$T$ phase exhibits different symmetry, as shown in the left column of Fig.\ref{fig:scenario}.
In scenario 1, the in-plane component $J_y$ forms an all-in/all-out configuration, whereas in scenario 2, $J_x$ forms a vortex-like configuration.
These configurations correspond to different magnetic space groups: $P32'1$ (no. 150.27) in scenario 1 and $P31m'$ (no. 157.55) in scenario 2.
This difference enables the determination of the OP through a detailed investigation of the magnetic structure in the low-$T$ phase. 
A recent theoretical study suggests that it is possible to distinguish between the $O_{yz}$ and $O_{zx}$ states using Sn-NMR under a magnetic field at a specific angle \cite{Kusunose2024}.
Another recent theoretical study has predicted the emergence of magnetoelectric effects depending on the $O_{yz}$ and $O_{zx}$ states in the kagome lattice system \cite{Ishitobi2023}.

The AFQ order that preserves threefold rotational symmetry in the kagome structure implies frustration among quadrupoles in URhSn.
In addition, the interplay between frustration and characteristic symmetries, including chirality and polarity, often gives rise to another class of nontrivial physical properties, such as cross-correlation phenomena \cite{Johnson2013, Mito2022}, anomalous magnetotransport \cite{Nakatsuji2015,Kakihana2018,Ishizuka2018}, and topological magnetic order \cite{Muhlbauer2009,Kaneko2019,Matsumura2017}.
Further investigation including cross-correlation properties in URhSn could provide new insights into the role of AFQ order.

In conclusion, the present RXS experiment unveiled a 5\textit{f}-electronic order associated with the successive phase transition of URhSn.
The nature of the high-$T$ phase was revealed to be the AFQ order with $\bm{q} = 0$.
The polarization dependence of the RXS signal in two configurations together with the group theory narrowed down the possible OPs to two candidates, $O_{yz}$ or $O_{zx}$.
Subsequently, the ground state below 16 K is now revised; a ferromagnetic phase reported earlier is indeed a canted antiferromagnetic phase owing to underlying AFQ order.
The present study reveals that in addition to crystal and spin, quadrupole degrees of freedom can also impart chirality and/or polarity to the system. 
Further microscopic investigations, including neutron diffraction, NMR, and elastic property measurements, are currently underway to determine the specific quadrupole OP and the resulting magnetic structure.

\begin{acknowledgments}
The authors are grateful to T. Ishitobi, K. Hattori, Y. Tokunaga, K. Tsuchida, T. Yanagisawa, H. Kusunose, and H. Harima for valuable discussions and comments.
The authors would also like to thank R. Murata and Y. Kaneko for assistance with the experiments.
This work was supported by JSPS KAKENHI Grants Nos. JP23H04867, JP23H04870, JP21K14644, and JP20H01864.
The synchrotron radiation experiments were performed at Photon Factory with the approval of Photon Factory Program Advisory Committee (Proposal Nos. 2020G072 and 2022G109). 
\end{acknowledgments}
\nocite{Suppliment} 
\bibliography{18110.bib}

\end{document}